**Reply to comments-on**

**"Brilliant source of 19.2 attosecond soft X-ray pulses below the atomic unit of time"**

**by Han (arXiv:2510.17949)**

Fernando Ardana-Lamas[1,†], Seth L. Cousin[1,†], Juliette Lignieres[1], Jens Biegert[1,2,*]

[1] ICFO - Institut de Ciencies Fotoniques, The Barcelona Institute of Science and Technology, 08860 Castelldefels (Barcelona), Spain.
[2] ICREA, Pg. Lluís Companys 23, 08010 Barcelona, Spain.
[*] Correspondence to: jens.biegert@icfo.eu
(11.November 2025)

Abstract:

We recently reported a refined analysis of a previously conducted soft X-ray (SXR) attosecond streaking measurement [1], employing the Variational Phase Gradient Temporal Analysis (VPGTA) retrieval algorithm [2]. This re-evaluation, prompted by new methodological insights, revealed a 19.2 attosecond pulse—consistent with the expectations and estimates of our earlier work [3]. Shortly thereafter, a comment by M. Han [4] challenged our findings, citing "physical and technical issues concerning the original experiment and the new characterization," including purported contributions from Auger electrons, unfiltered low-energy harmonics, and uncompensated intrinsic chirp. These concerns, however, stem from a misinterpretation of the experimental regime, selective citation of context, and disregard for well-established results in the peer-reviewed literature. In this reply, we address and clarify these points, demonstrating that the issues raised are not relevant under our experimental conditions. Given the highly specialized nature of attosecond generation and metrology in the soft-X-ray water-window regime, we consider it essential to clarify these aspects of different regimes and reaffirm the validity of our conclusions.

We recently published new results from a previously conducted streaking measurement of an SXR attosecond pulse [1] using the VPGTA [2] retrieval algorithm. As per scientific practice, we felt compelled to do so because new insights were available, and we found the measurement revealed a 19.2 attosecond pulse, confirming estimates from our earlier work [3]. However, we recently learned of an arXiv comment on our work by M. Han [4]. He contests our result and voices "(…) physical and technical issues concerning the original experiment and the new characterization. Specifically, without accounting for the contribution of Auger electrons or employing filters to remove low-energy harmonics and compensate for the intrinsic pulse chirp."

Here, we address M. Han's accusations since they are based on misleading claims, confused generation regimes, and ignoring peer-reviewed published literature. We are concerned about M. Han's confusing comments, as only a few people are intimately familiar with the differences and pitfalls of attosecond generation and metrology in the soft X-ray water window regime.

We note that M. Han never contacted us to discuss his concerns and that he is also the last author of an arXiv article [5] in which he claims a 25-attosecond pulse covering the carbon K-shell edge and a three-order-of-magnitude increase in harmonic flux over previous published values.

In the following, we address his comments to our publication.

**1. Photoelectrons from krypton valence orbitals (4s and 4p) ionized by low-energy harmonics:**

*Han: " (…) did not measure the harmonic spectrum below 150 eV due to the limited working range of the grating (Hitachi, 2,400 lines per mm according to [9]) and employed no filters to suppress these low-energy components. However, harmonics below 100 eV can generate a significant number of photoelectrons through ionization of the valence orbitals, which would spectrally overlap with the photoelectrons from the 3d inner-shell orbital ionized by higher-energy harmonics (>100 eV). As shown in Fig. 1 of [2], the photoionization cross section of the krypton valence orbitals is substantially larger than that of the 3d orbital in this low-energy range. Therefore, attributing the measured photoelectrons solely to ionization of the 3d orbital by harmonics above 100 eV—without showing the HHG spectrum and accounting for their contribution in retrieval—is highly problematic and undermines the reliability of their pulse retrieval."*

This interpretation overlooks the defining features of high-harmonic generation under high-pressure phase-matching conditions [6-10], also known as non-adiabatic phase-matching [6]. As firmly established in the literature, at high pressures, the generation regime bears little resemblance to that encountered with near-infrared drivers at lower pressures. A decisive difference is the strong reabsorption of lower-energy harmonics within the medium. Using Henke data [12], even a 1 mm path at 1 bar results in complete absorption of photon energies relevant for our measurement below 150 eV—conditions far less stringent than those of our experiment (3 mm, 3.5 bar); see Fig. 1. This is consistent with measurements.

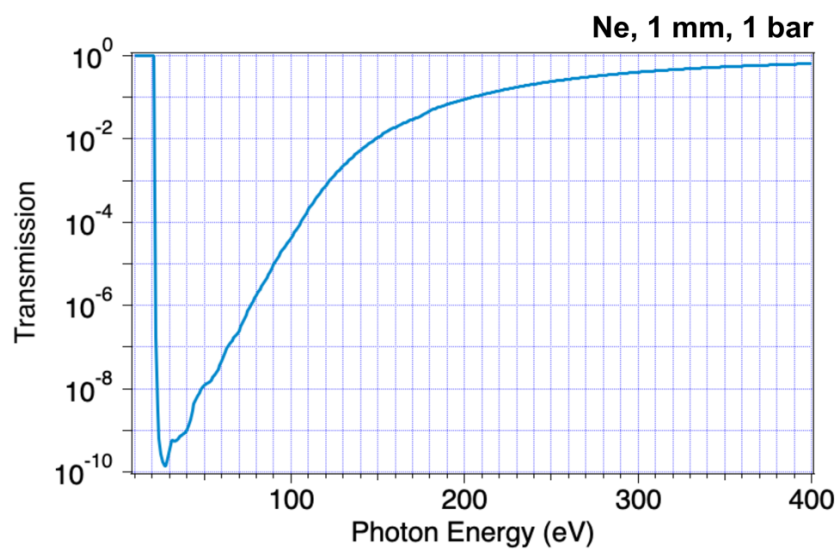

Fig. 1: Transmission as a function of photon energy through a 1 mm long path of neon at 1 bar. Based on Henke's data [12].

The discussion of possible contributions from krypton valence-orbital ionization does not address the relevant photoionization dynamics. The 3d orbital can only be accessed with photon energies above 94 eV, where its cross section exceeds that of the valence orbitals by up to 1.5 orders of magnitude.

**2. Intrinsic positive chirp of the attosecond pulses generated by HHG**:

*Han: "Without employing any filters, Ref. [1] reports a nearly chirp-free pulse in Fig. 3b, without specifying the sign of the small residual chirp. However, this result is inconsistent with the combined contribution of the intrinsic attochirp—estimated by the authors themselves to range from 1654 as² to 6172 as²—and the propagation-induced chirp of –101.3 as² predicted by their fluid dynamics simulation (see Fig. 5g), which is only very briefly introduced. A more robust and convincing approach would be to use spectral filters and directly measure the filter-induced chirp from the corresponding streaking traces."*

This comment quotes isolated numbers from our previous publications without context or physical meaning. As discussed in detail in our 2015 paper [3], the semiclassical estimate of single-atom propagation neglects macroscopic propagation and phase-matching effects. Such a quantity, therefore, cannot be compared to the effective chirp of the experimentally observed pulse, which inherently includes propagation-induced effects and non-adiabatic plasma dynamics.

Moreover, the conditions under which high-harmonic generation is driven in our work are a high-ionization regime, where the ionization fraction rises by orders of magnitude within a fraction of an optical cycle, and both the temporal phase and the macroscopic phase matching evolve. This behaviour was already discussed in Ref. [6,8-10]. We reiterate the early work of Tempea et al. in 2000 [6] discussing non-adiabatic phasematching, a first investigation of M.-C. Chen et al. [7] showed a transition from attosecond pulse train generation to isolated attosecond pulse generation despite using a multicycle pulse when increasing pressure, our work, which discussed phase-matching at high pressure and under strong ionization [8,9], and the work of A.S. Johnson, in which they discuss phasematching for SXR generation [10].

The suggestion to employ spectral filters as diagnostics by changing chirp is misplaced: while such filters can impose appreciable chirp in the XUV regime, they are of little use in the SXR water window photon energy range; this is shown in our publication [1] in Fig. 4. We note that the work of Han [5] shows in Fig. 2D spectra whose high photon energy part just about crosses the carbon edge at 284 eV. Still, theyir data does not exhibit any absorption edge modulation.

**3. Discrepancy between measured and retrieved streaking traces**:

*Han: "The measured streaking trace shown in Fig. 2c of [1] clearly exhibits sub-cycle asymmetries, which indicate a chirped attosecond pulse. However, the retrieved trace presented in Fig. 2b of [1] appears artificially "perfect," lacking any such asymmetry. This discrepancy*

*between the measured and retrieved streaking traces raises serious concerns about the reliability of their pulse characterization. Moreover, the streaking trace in Fig. 2c of [1] appears very similar to Fig. 4 of [2], which was later retrieved using the autocorrelation representation in [6] to yield a pulse duration of ~165 as. In contrast, the current retrieval result is about eight times shorter, raising questions about how such a significant discrepancy arises."*

This statement conflates independent datasets and misinterprets the content of our figures. Both the measured and retrieved traces are shown explicitly in our publication, and their agreement—including residual sub-cycle structure—is evident. The retrieved pulse is not claimed to be transform-limited; rather, its small residual chirp is clearly displayed and discussed in Fig. 3a of Ref. [1]. The retrieval, therefore, accurately reflects the experimental data within the limits of the measurement.

The reference to Ref. [12] is misplaced, as that work is unrelated to ours. The cited "165 as" result refers to an autocorrelation retrieval method not applied in our study. The inference that the two datasets should yield identical pulse durations is thus unfounded. Our analysis follows the same retrieval formalism established in Ref. [2] and widely validated in the community. Han's method shares the same foundational principles, differing only in computational implementation.

Han further states in Ref. [4] " (…) *the retrieved trace presented in Fig. 2b of [1] appears artificially "perfect," lacking any such asymmetry. This discrepancy between the measured and retrieved streaking traces raises serious concerns about the reliability of their pulse characterization.*" A simple lineout of the streaking data shown below in Fig. 2 shows that this statement is wrong.

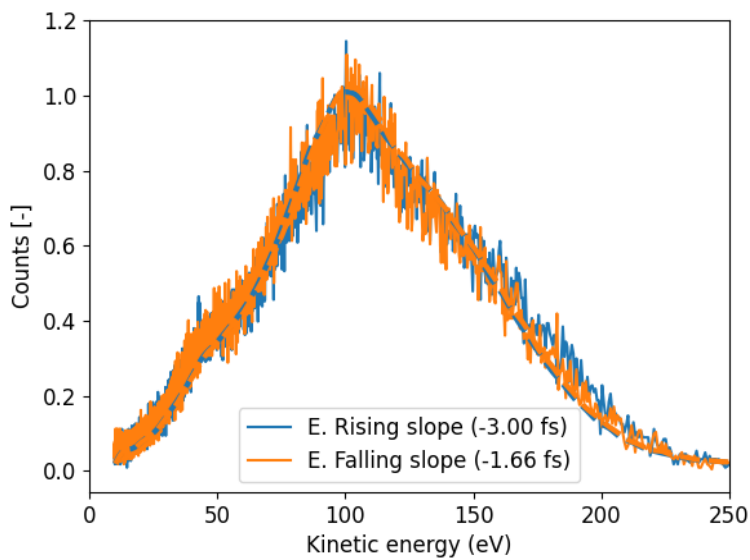

Fig. 2: Lineout from data [1]. Experimental photoelectron spectra at the two different slopes of electric field where "asymmetries" are expected to be larger. Dashed lines, represents the corresponding retrieve spectra.

Within this framework, both the measured and retrieved streaking traces are fully consistent, and no discrepancy of the kind suggested arises.

**4. Non-negligible contribution of Auger electrons for atomic inner-shell ionization:**

*Han: "We have also conducted attosecond streaking experiments on krypton atoms [10] using isolated attosecond pulses with a central photon energy of ~130 eV (bandwidth ~20 eV), where inner-shell ionization resulted in a significant yield of Auger electrons. These electrons are notably absent in the experimental data and retrieval process presented in [1]. As shown in Figure 1, in the energy range of 30–60 eV the Auger electron yield is comparable to that of the 3d photoelectrons, agreeing with the measurement with synchrotrons [11]. Importantly, this electron energy range was used for retrieving the attosecond pulse in Fig. 3b of [1]. Ignoring the contribution of Auger electrons in the pulse retrieval process raises serious concerns about the accuracy and validity of their characterization—additionally, Refs. [1,2] did not show the energy calibration of the time-of-flight (TOF) electron spectrometer—a crucial step, particularly for high-energy electrons [7]. This calibration is typically performed using the energies of Auger electrons."*

Our PRX [3] publication extensively discusses possible Auger-electron contributions, including quantitative analysis of branching ratios, measurement of krypton ion yield, and experimental validation. The outcome is unambiguous: under our experimental conditions, such contributions are negligible and can safely be disregarded.

By contrast, the measurement presented by Han [5] lacks the essential information required to assess this question. The corresponding HHG photon spectrum is not provided, and Ref. 10 cited in their work [4] as its source merely states that it will be published elsewhere. Without this key information, any claim about which krypton shells contribute to the reported signal remains speculative. It is therefore inappropriate to dismiss our results based on such incomplete data.

Finally, a direct comparison between our measurement, —where all relevant data and calibration details are explicitly reported—and Han's shows that the photon-energy range in which Auger processes could contribute is entirely outside the regime probed in our experiment. Figure 2 shows a direct scaled comparison between Fig. 1 of Ref. [4] and our data. Therefore, the suggestion that Auger electrons influence our results is unsupported both by physical reasoning and available evidence.

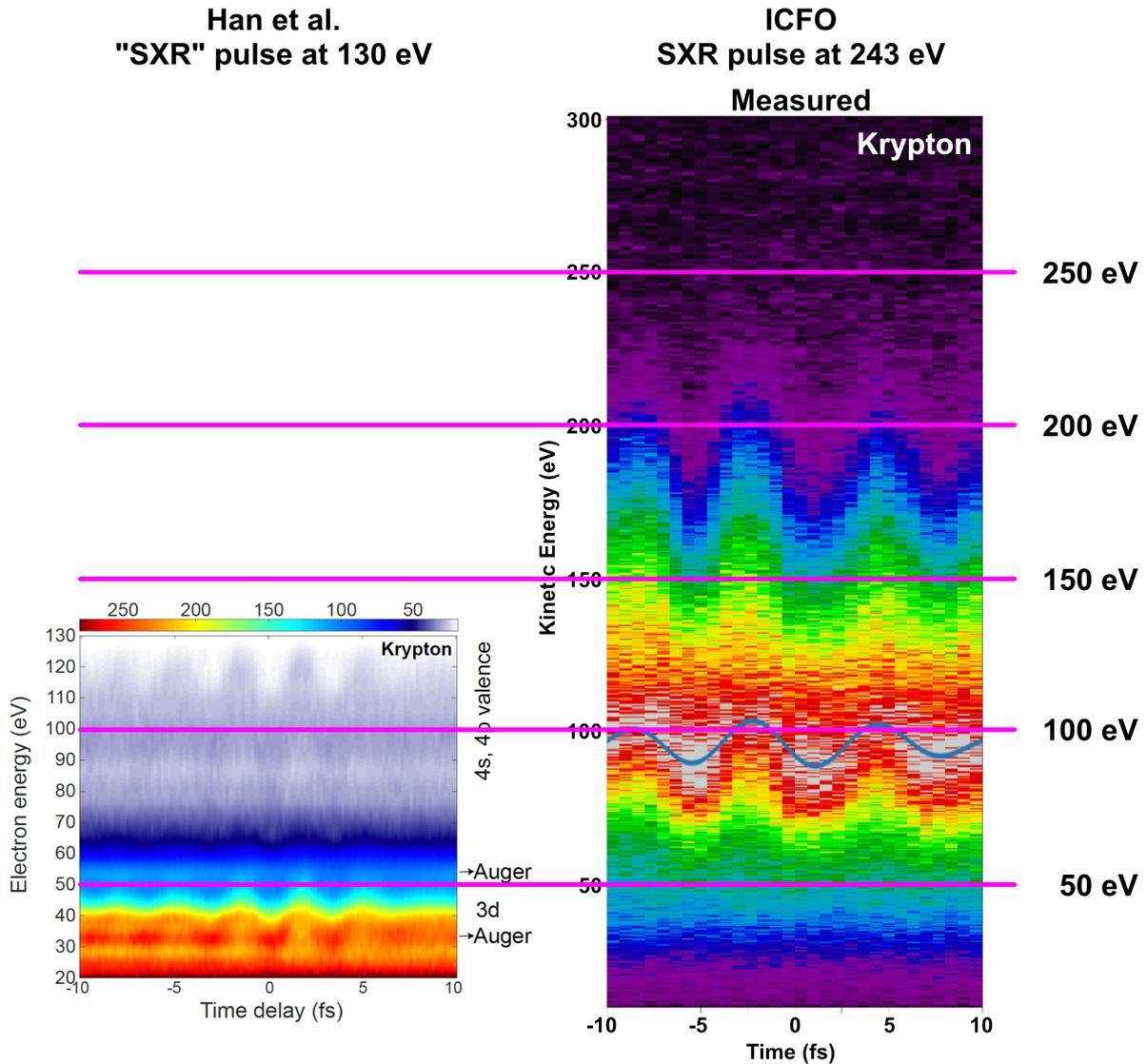

Fig. 3: Comparison of the data from Han's comment [4] (Fig. 1) and our data [1]. Clearly shown is the very different photon energy range that leads to an entirely different electron energy distribution without relevant contribution of Auger electrons in accordance with Ref. [3].

**5. Photon flux calibration and discrepancy:**

Han: *"Ref. [1] claims an overall photon flux of 4.8×10$^{10}$ photons per second, but provides no measurement details regarding the calibration procedure. Accurate photon flux measurement requires the use of a filter to remove the residual driving laser field fully. However, as mentioned above, that work did not use any filter to remove the driving field. Notably, the reported photon flux in [1] is almost five orders of magnitude compared to the previously published value of 5.6×10$^5$ photons per second from 284 to 350 eV on target under similar experimental conditions in [2]."*

The discrepancy described by Han is based on a misunderstanding of how photon flux is defined and reported in our publications. We have consistently provided photon numbers for specific, well-defined spectral ranges relevant to applications—namely, the photon flux at the carbon K-shell absorption edge and the integrated flux over the whole harmonic spectrum. These quantities are distinct and serve different purposes. It is therefore unsurprising, and entirely consistent, that their numerical values differ by several orders of magnitude. Also, the measurements were taken at different times under slightly different setups but the overall numbers are consistent. We list them here with the first column providing the year of publication with reference, the second column specifying whether the reported value is on target where experiments are conducted or at the generation point, the third column illustrated whether the value are given for the range from an absorption edge to the cutoff (a), at a specified bandwidth at the absorption edge (b), or integrated over the entire spectrum (c). The last two columns provide the photon energy values and the reported photon flux or source brilliance and the generation medium.

| | | | | |
|---|---|---|---|---|
| 2014 Ref. [13] | HHG | (b) | 284 eV | $1.85 \times 10^7$ photons/s/1%; neon |
| 2014 Ref. [13] | HHG | (b) | 284 eV | $4.3 \times 10^{15}$ photons/s/mm$^2$/strad/10%; neon |
| 2016 Ref. [8] | HHG | (a) | 284 – 550 eV | $7.3 \times 10^7$ photons/s; neon |
| 2016 Ref. [8] | HHG | (b) | 284 eV | $2.8 \times 10^7$ photons/s/10%; neon |
| 2016 Ref. [8] | HHG | (b) | 284 eV | $1.8 \times 10^6$ photons/s/10%; helium |
| 2017 Ref. [3] | Target | (a) | 284 – 350 eV | $5.6 \times 10^5$ photons/s; neon |
| 2017 Ref. [1] | HHG | (c) | 200 – 380 eV | $4.8 \times 10^{10}$ photons/s; neon |
| 2017 Ref. [1] | HHG | (b) | 284 eV | $4.1 \times 10^9$ photons/s/10%; neon |

These figures are mutually consistent and correspond precisely to the different integration ranges stated in each case.

By contrast, Han's "five orders of magnitude discrepancy" [4] arises from comparing quantities defined over fundamentally different bandwidths, effectively conflating localized photon flux with the broadband total. Furthermore, his assertion regarding filter use is misplaced. Finally, the claim about operating conditions betrays a misunderstanding of the relevant photon-energy regime. Performing streaking measurements in the true soft-X-ray water window requires photon energies well above those at which helium can be used as a streaking gas, or orders of magnitude higher photon flux than presently provided by any harmonic generation source —conditions not met in Han's experiment either.

Summary:

The comment by M. Han reflects a series of misinterpretations of our experimental conditions and methodology, leading to conclusions that do not apply to the soft-X-ray attosecond generation regime investigated in our work. All relevant aspects of our experiment—including harmonic generation, chirp compensation, streaking analysis, and photon-flux calibration—are comprehensively documented and consistent with established results in the field. While such public discussion can be valuable to the community, it must be based on an accurate understanding and fair representation of the underlying science. Our clarification here reaffirms the integrity of our data, the validity of our retrieval, and the robustness of the conclusions presented in our original publication.